\documentclass{PoS}
\usepackage{amsmath,amssymb}
\usepackage{graphics}

\title{The Upper Energy Limit of HBChPT in Pion Photoproduction}

\ShortTitle{The Upper Energy Limit of  HBChPT in Pion Photoproduction }

\author{\speaker{C\'esar Fern\'andez-Ram\'irez}\\
        Grupo de F\'{\i}sica Nuclear, 
Departamento de F\'{\i}sica At\'omica, Molecular y Nuclear,
 Facultad de Ciencias F\'{\i}sicas, Universidad Complutense de Madrid, CEI Moncloa,
 Avda. Complutense s/n, E-28040 Madrid, Spain\\
        E-mail: \email{cefera@gmail.com}}

\abstract{We assess the energy limit up to which Heavy Baryon Chiral Perturbation Theory (HBChPT)
can be applied to the process of neutral pion photoproduction from the proton 
by analyzing the latest data collected by the MAMI/A2 collaboration at Mainz.
The high-quality differential cross section and the photon asymmetry measured data allow to
test the theory at a level not previously achieved.
We find that, within the current experimental status, the agreement between theory and experiment is excellent
up to $\sim$170 MeV. Above this energy HBChPT fails to provide high-quality fits while an empirical parametrization of the multipoles still provides an excellent description of the data.}

\FullConference{The 7th International Workshop on Chiral Dynamics,\\
		August 6 -10, 2012\\
		Jefferson Lab, Newport News, Virginia, USA}

\begin{document}

\section{Introduction}
The spontaneous breaking of chiral symmetry in Quantum Chromodynamics (QCD) 
makes the $\pi$ meson appear as a pseudoscalar Nambu--Goldstone boson \cite{book}.
This has some dynamical consequences and
one of the most prominent is the softness of the
S-wave amplitude for the $\gamma N \rightarrow \pi^{0} N$ reaction in the near
threshold region, since it vanishes in the chiral limit \cite{CHPT}. 
As a consequence of this softness and the large P-wave amplitude contribution 
due to the early appearance of the $\Delta(1232)$ \cite{AB-Delta},
the S- and P-wave contributions
are comparable very close to threshold \cite{AB-fits} and even D waves matter \cite{FBD09}.
Hence, the accurate extraction of the S and P waves from pion photoproduction data becomes
an important issue in the study of chiral symmetry breaking and QCD.
Consequently, neutral pion photoproduction from the proton has constituted one of the most studied reactions 
to test chiral dynamics in the baryon sector, both from the experimental and theoretical perspectives. 
The A2 and CB-TAPS collaborations at MAMI (Mainz) have recently measured the differential cross section and 
the photon beam asymmetry \cite{Hornidge} in the low energy region for neutral pion photoproduction 
with such precision that it is possible to extract the P-wave energy dependence and
to use the data to accurately test Heavy Baryon Chiral Perturbation Theory (HBChPT)
and assess the energy range where the theory is accurate.

\section{Structure of the Observables}
The differential cross section and the photon beam asymmetry ($\Sigma$) can be expressed in terms of the
electromagnetic responses \cite{FBD_PRC09}:

\begin{eqnarray}
\frac{d \sigma }{d\Omega}
\left( s, \theta \right) & = & \frac{q}{k_\gamma} W_{T}\left( s, \theta \right) \label{eq:eq1} \\
\Sigma \left( s, \theta \right) &\equiv& \frac{\sigma_\perp - \sigma_\parallel}{\sigma_\perp + \sigma_\parallel} =-\frac{W_{S}\left( s, \theta \right)}{W_T\left( s, \theta \right) }
\sin^2 \theta \label{eq:eq2}
\end{eqnarray}
where $W_T$ and $W_S$ are the electromagnetic responses, 
$\theta$ is the center of mass scattering angle,
$k_{\gamma}$ the center of mass photon energy, 
$q$ the pion momentum in the center of mass, and
$s$ the squared invariant mass.
The responses  $W_T$ and $W_{S}$
are defined in term of the electromagnetic multipoles:
\begin{equation}
W_{T} =T_0\left( s \right) + T_1\left( s \right) \mathcal{P}_1\left( \theta \right)  
+ T_2\left( s \right) \mathcal{P}_2\left( \theta \right) + \dots \label{eq:wt}
\end{equation}
\begin{equation}
W_{S} = S_0 \left( s \right) +S_1 \left( s \right) \mathcal{P}_1\left(  \theta \right)+ \dots \label{eq:wtt}
\end{equation}
where $P_j \left( \theta \right)$ are the Legendre polynomials in terms 
of $\cos \theta$, the dots stand for negligible corrections, and
\begin{eqnarray}
T_n \left( s \right)&=&\sum_{ij} \text{Re} \{ \: \mathcal{M}^*_i \left( s \right) \: 
T_n^{ij} \: \mathcal{M}_j \left( s \right) \: \} \label{eq:Tn} \\
S_n \left( s \right)&=&\sum_{ij} \text{Re} \{ \: \mathcal{M}^*_i \left( s \right) \: 
S_n^{ij} \: \mathcal{M}_j \left( s \right) \: \} 
\end{eqnarray}
where up to D waves
$\mathcal{M}_j \left( s \right) =E_{0+}$, $E_{1+}$, $E_{2+}$, 
$E_{2-}$, $M_{1+}$, $M_{1-}$, $M_{2+}$, $M_{2-}$.
The detailed analysis of the partial wave structure of the observables and the 
coefficients $T_n^{ij}$ and $S_n^{ij}$ can be found in \cite{FBD_PRC09}.

\section{Results}

\begin{figure}
\begin{center}
\rotatebox{-90}{\scalebox{0.4}[0.4]{\includegraphics{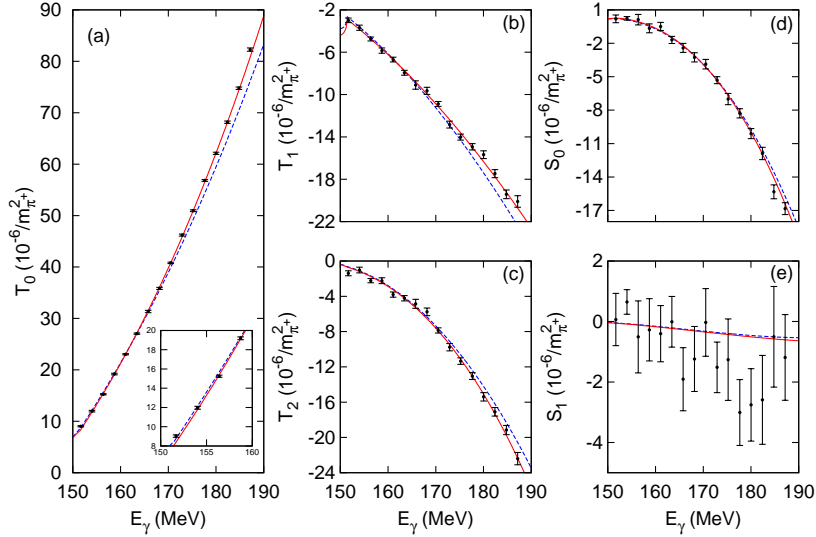}}}
\caption{Coefficients of the Legendre polynomial expansion of the observables. Data from \cite{Hornidge}. The solid red line stands for the empirical fit and the dashed blue line for the HBChPT fit. The inset in figure (a) provides a better look at $T_0$ in the lowest energy region.} \label{fig:Ts}
\end{center}
\end{figure}

From the differential cross sections and the photon beam asymmetries measured in \cite{Hornidge} 
it is possible to extract the $T_j$ and $S_j$ coefficients. In Figure \ref{fig:Ts} we show them. 
The extracted $S_1$ is compatible with zero and provides no additional information \cite{FB13}. 
Hence, we are able to extract four single-energy quantities -- i.e. $T_0$, $T_1$, $T_2$, and $S_0$ --, 
from the data, what implies that we can only extract four single-energy multipoles, in our case 
$\text{Re} E_{0+}$, $\text{Re} E_{1+}$, $\text{Re} M_{1+}$ and $\text{Re} M_{1-}$, 
assuming that the imaginary part of the P waves is zero, $\text{Im} E_{0+}$ is fixed through unitarity and 
D waves are fixed by the Born terms. The single energy multipoles can be found in \cite{Hornidge,FB13}.

Together with the single-energy multipoles one can fit the experimental data to different theoretical approaches that describe the multipoles. We employ three approaches:
1) HBChPT calculations to $O(q^4)$ \cite{FBD09,HBCHPT} with the five empirical
low-energy constants brought up to date by fitting these
data \cite{FB13}; 2) relativistic ChPT calculations (also to $O(q^4)$)
which as well have five low-energy constants fit to these
data~\cite{Hilt2012}; and 3) an empirical fit \cite{Hornidge,FBD_PRC09}:
\begin{eqnarray}
E_{0+} &=& E_{0+}^{(0)} + E_{0+}^{(1)} \frac{\omega-m_{\pi^0}}{m_{\pi^+}}
+ i\beta \frac{q_{\pi^+}}{m_{\pi^+} } \: ,  \label{eq:Swave} \\
P_j \slash q &=&  \frac{P_{j}^{(0)} }{m_{\pi^+}}+  P_j^{(1)} 
\frac{\omega-m_{\pi^0}}{m^2_{\pi^+}}  \: \: \text{;} \: \: j=1,2,3 \label{eq:Pwave}
\end{eqnarray}
where $E_{0+}^{(0)}$, $E_{0+}^{(1)}$, $P_{1}^{(0)}$, $P_{1}^{(1)}$, $P_{2}^{(0)}$, 
$P_{2}^{(1)}$, $P_{3}^{(0)}$, and $P_{3}^{(1)}$ are free parameters 
that will be fitted to the experimental data, $\omega$ is the pion energy in the center of mass 
and $\beta$ is fixed through unitarity \cite{FB13}. 
We note that chiral symmetry is not imposed in this approach 
and that $P_j$ partial waves are related to standard electromagnetic multipoles through
\begin{eqnarray}
E_{1+}  &=& \left( P_1 + P_2 \right)/6  \\
M_{1+} &=&  \left( P_1 - P_2 \right)/6 + P_3/3   \\
M_{1-}  &=& \left( P_3 + P_2 - P_1 \right)/3 
\end{eqnarray}

\begin{figure}
\begin{center}
\rotatebox{-90}{\scalebox{0.35}[0.35]{\includegraphics{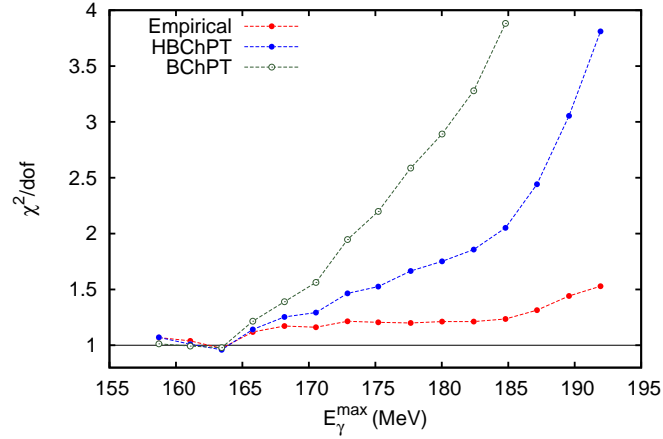}}}
\caption{$\chi^2\slash$dof energy dependence for the
empirical (full red circles) \cite{Hornidge}, 
HBChPT (full blue circles) \cite{FB13}, and
BChPT (empty green circles) \cite{Hilt2012}
fits. Each point represents a separate fit and the connecting lines are drawn to guide the eye. 
The points are plotted at the central energy of each bin, although the calculations 
take the energy variation inside of each bin into account. 
The value $\chi^2\slash$dof$=1$ is highlighted with a solid line.} \label{fig:chi2}
\end{center}
\end{figure}

\begin{figure}
\begin{center}
\begin{tabular}{cc}
\rotatebox{-90}{\scalebox{0.27}[0.27]{\includegraphics{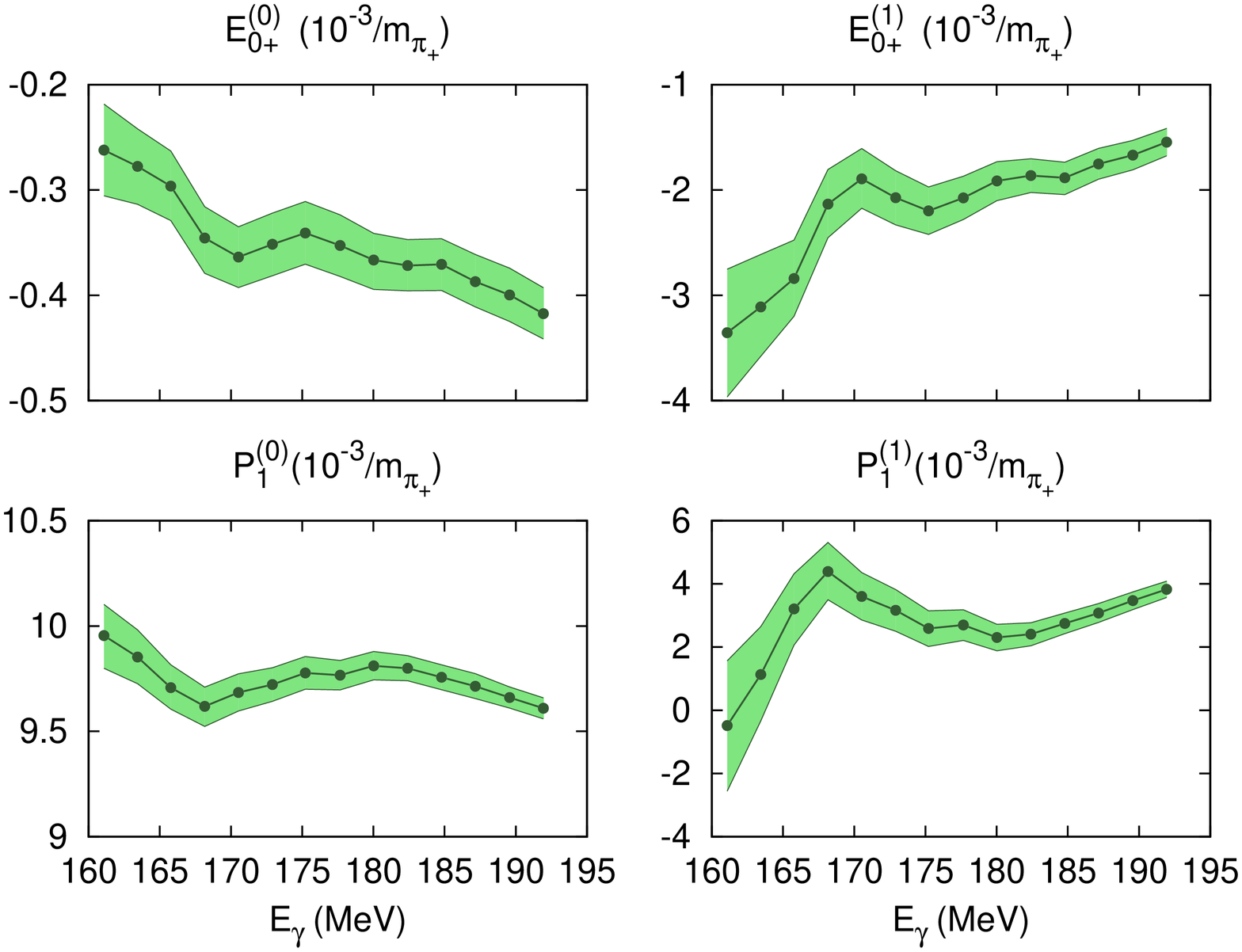}}} &
\rotatebox{-90}{\scalebox{0.27}[0.27]{\includegraphics{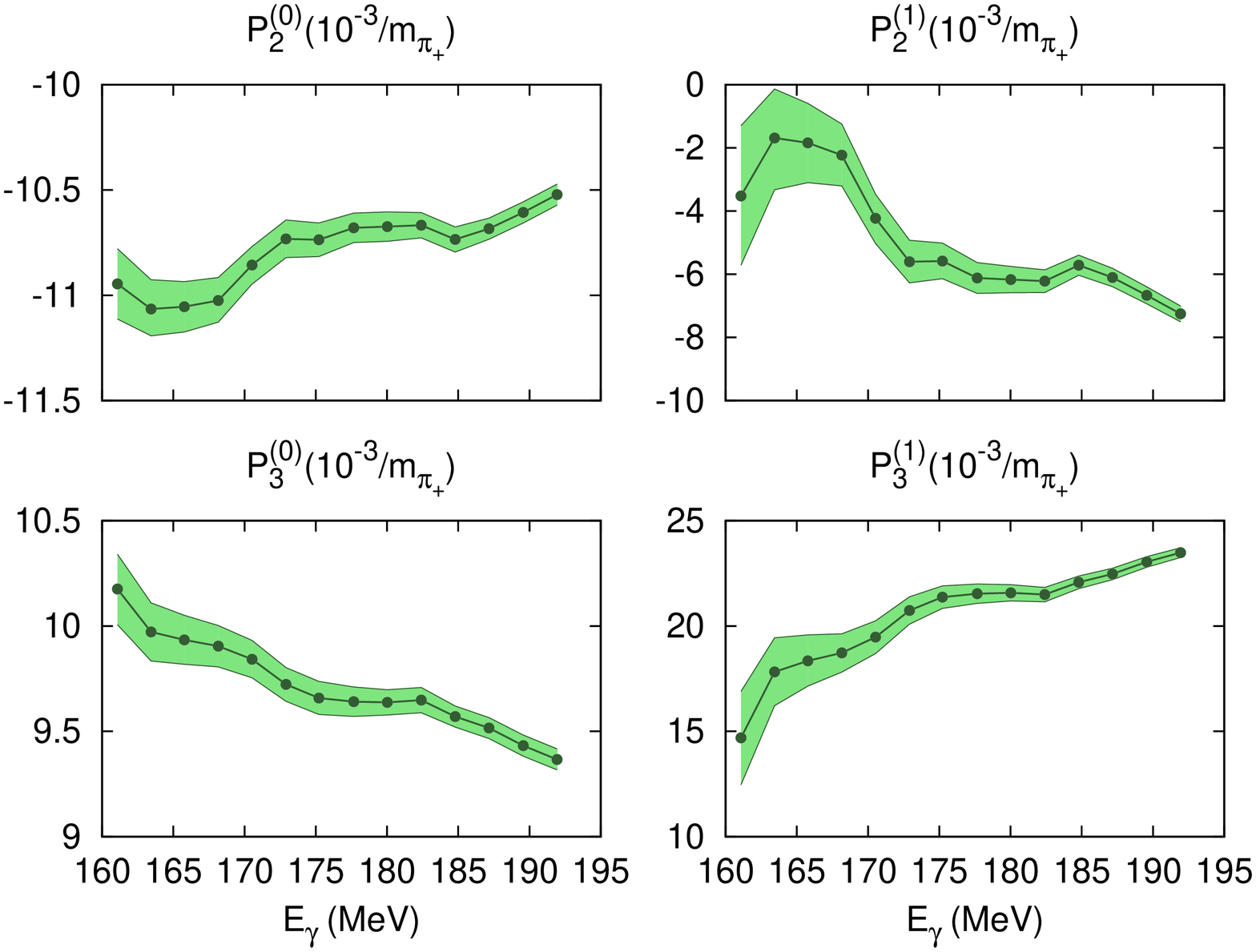}}}
\end{tabular}
\caption{Upper energy (fit) dependence of the parameters for the empirical fit. 
Error bands are computed at the $\chi^2_\text{min}+1$ level as described in \cite{Hornidge,FB13}.} 
\label{fig:LECs}
\end{center}
\end{figure}

We perform fits to the experimental data in  \cite{Hornidge} up to different maximum 
photon energies $E_\gamma^{\text{max}}$ within the range  $\left[ 158.72,191.94 \right]$ MeV 
and compute the $\chi^2\slash$dof. 
The amount of data employed in each fit depends on up to what energy we are fitting,
--- i.e. for our lowest-energy fit ($E_\gamma^{\text{max}}=158.72$ MeV) 
we employ $100$ experimental data 
and for our highest-energy fit ($E_\gamma^{\text{max}}=191.94$ MeV) 
we employ $514$ experimental data.
Systematics are not included in the $\chi^2$ 
and this uncertainty can amount up to 4\% in the differential cross section 
and 5\% in the photon asymmetry. 
Figure \ref{fig:chi2} shows the $\chi^2\slash$dof for every fit performed versus the upper energy 
$E_\gamma^{\text{max}}$ of the fit as well as the number of data.
It is shown that up to $\sim$170 MeV all the fits are equally good providing 
very low $\chi^2\slash$dof.
Above 170 MeV the trend is different; while the empirical fit remains with a good and stable 
$\chi^2\slash$dof, both the HBChPT and the BChPT start rising, a trend that shows clearly 
how the theory fails to reproduce the experimental data above that energy.
The empirical fit provides good agreement with the data
up to $\sim$185 MeV where the imaginary parts of the P waves start to be important.
The parameters of the empirical fits are approximately constant on $E_\gamma^{\text{max}}$
as we is show in Figure \ref{fig:LECs}.
The same graphs for the HBChPT low energy constants can be found in \cite{FB13}.

\begin{figure}
\begin{center}
\begin{tabular}{cc}
\rotatebox{-90}{\scalebox{0.28}[0.28]{\includegraphics{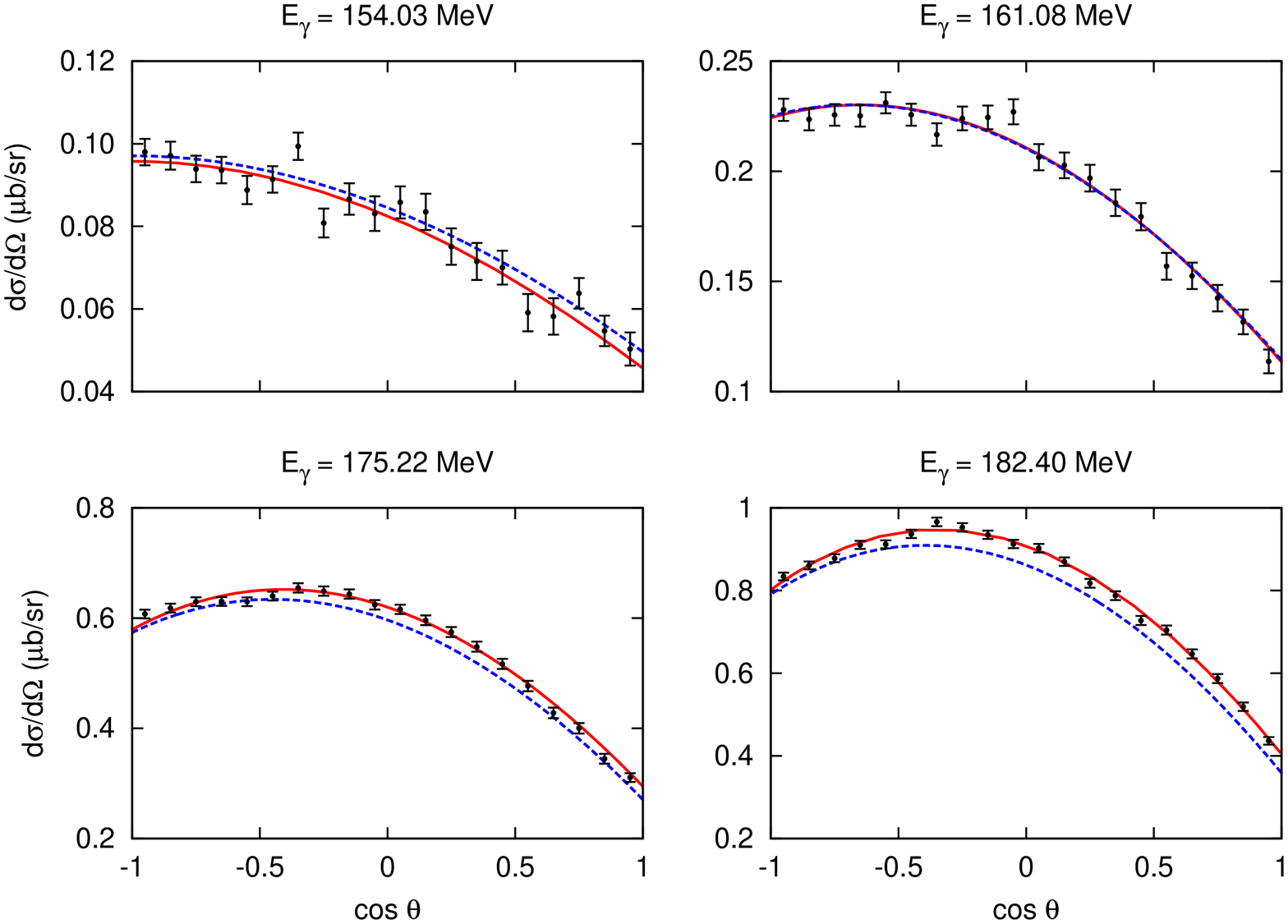}}} &
\rotatebox{-90}{\scalebox{0.28}[0.28]{\includegraphics{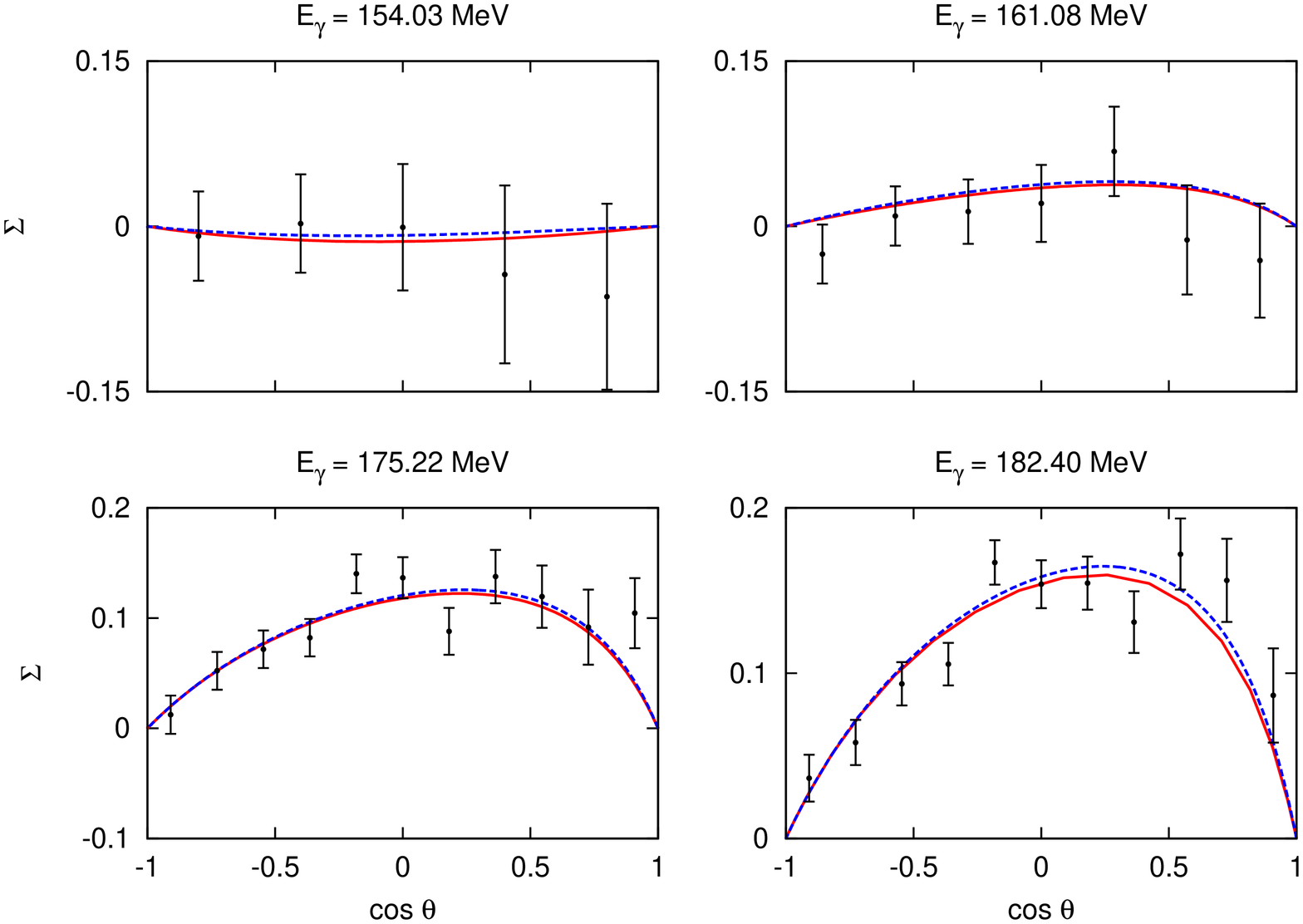}}}
\end{tabular}
\caption{Differential cross section and photon beam asymmetry at different energies. Data from \cite{Hornidge}. Solid red: empirical; Dashed blue: HBChPT.} \label{fig:observables}
\end{center}
\end{figure}

In Figure \ref{fig:observables} we compare the empirical and HBChPT calculations 
for the differential cross section and the photon beam asymmetry at different energies.
For the differential cross section it is clear that above 170 MeV the HBChPT approach
does not provide a good description of the data, however, for the photon beam asymmetry
the agreement is good in the whole energy range.

\section{Conclusions}
The main results we have obtained can be summarized in:
\begin{enumerate}
\item The high-quality experimental data gathered by the A2 and CB-TAPS collaborations
at MAMI allow to obtain the electromagnetic multipoles and their energy dependence 
to the best precision ever, what allows to accurately test chiral symmetry and
the range of application of HBChPT.
\item We can establish a clear upper-energy limit of validity for HBChPT. We have found this limit
 to be $\sim$170 MeV of photon energy in the laboratory frame. 
\item We do not think that calculating higher orders in HBChPT will help extending the energy
range of application of the theory because data are fairly 
well reproduced by the empirical fit up to 185 MeV which expands up to a lower order in pion energy. 
\item Further improvement is necessary in the theory if ChPT is to be applied above 170 MeV.
The relativistic calculation does not provide better agreement with data 
than the HBChPT approach \cite{Hornidge,Hilt2012} what suggests
the inclusion of the $\Delta$(1232) in the calculation  \cite{DeltaBCHPT} .
\end{enumerate}

\acknowledgments
The work reported here was carried out in collaboration with A. M. Bernstein.
The author thanks the A2 and CB-TAPS collaborations for making available 
the experimental data prior to publication.
C.F.R. is supported by \textit{Juan de la Cierva}
programme of  Spanish Ministry of Economy and Competitiveness and his
research has been conducted with support by
Spanish Ministry of Economy and Competitiveness 
grant FIS2009-11621-C02-01,
the Moncloa Campus of International Excellence (CEI Moncloa),
and by CPAN,  CSPD-2007-00042 Ingenio2010. 
We gratefully acknowledge Jefferson Lab and the organizers for their hospitality.

\end{document}